# Actively tunable plasmonic lens for subwavelength imaging at different wavelengths

**Beibei Zeng[1,3], Haofei Shi[2] and Xiangang Luo[1]**

1. *State Key Laboratory of Optical Technologies for Microfabrication, Institute of Optics and Electronics, Chinese Academy of Science, Chengdu 610209, China*

2. *Department of Electrical Engineering and Computer Science, The University of Michigan, Ann Arbor, Michigan 48109, USA*

3. *Present address: Department of Electrical and Computer Engineering, Lehigh University, Bethlehem, PA 18015, USA*

E-mail: **bez210@lehigh.edu** and **lxg@ioe.ac.cn**

**Abstract:** A type of tunable plasmonic lens with nanoslits is proposed for subwavelength imaging in the far field at different wavelengths. The nanoslits array in the plasmonic lens, which have constant depths but varying widths, could generate desired optical phase retardations based on the particular propagation property of the Surface Plasmon Polaritons (SPPs) in the metal-dielectric-metal (MDM) slit waveguides. We theoretically and numerically demonstrate the tunability of a single plasmonic lens for subwavelength imaging (full width at half maximum, $0.37\lambda \sim 0.47\lambda$) by adjusting the surrounding dielectric fluid, thereby realizing the compact in-plane tunable plasmonic lens. This work provides a novel approach for developing integrative tunable plasmonic lens for a variety of lab-on-chip applications.

1. **Introduction**

Recently, the diffraction limit of conventional optics seems to have been overcome by the perfect lens proposed by J. B. Pendry using a slab of negative refractive index (NRI) media [1]. However, the difficulty of finding a homogenous NRI media restrains applications of the perfect lens. Considering the electrostatic approximation of the perfect lens, the superlens could also be used to realize sub-diffraction-limited imaging, only requiring negative permittivity that is available in natural metals at particular frequency ranges. Therefore, a lot of impressive works concerning superlenses have been achieved theoretically and experimentally [2-4]. However, the great shortcoming of the superlens is that the object and image should be confined in the near field, usually tens of nanometers away from the superlens [1, 5]. Therefore, an optical far-field superlens (FSL) device has been designed for imaging beyond the diffraction limit from far-field measurement [6].

On the other hand, as indicated at the end of Ref [7], a planar metallic slab with arrayed nanoslits of varying widths could be demonstrated theoretically [8] and experimentally [9] to focus light in the far field. And a planar plasmonic lens has been proposed to realize subwavelength imaging for arbitrary object and image distances [10]. Based on the optimum design of nanoslit waveguides in the metallic slab, the imaging process is achieved by manipulating the phase distribution of the optical field. Each nanoslit in the metallic slab is designed to transmit light with specific phase retardation, therefore arbitrary phase modulation on the wavefront could be realized. However, due to the dependence of permittivity of metallic materials on frequencies, applications would be limited because the plasmonic lens can only operate at a specific frequency when the positions and widths of the nanoslits are fixed.

Optical systems synthesized by using fluids are called optofluidics. Fluids have unique properties that cannot be found in solids and these properties can be used to design novel devices, such as oil-immersion microscopes [11], liquid mirrors for telescopes [12], liquid-crystal displays [13] and tunable liquid gradient refractive index lenses [14]. Integration and reconfiguration are the two main advantages of optofluidics. The second advantage of optofluidics means that one can easily change the optical properties of the devices by controlling the fluids [15]. Therefore, in this paper, we propose a tunable optofluidic planar plasmonic lens for subwavelength imaging at different incident wavelengths by simply varying the surrounding dielectric fluid.

## 2. Working principle

Fig. 1 is the schematic drawing of the optical imaging process for the tunable plasmonic lens. The plasmonic lens is a silver slab of thickness $d$ with nanoslits located symmetrically with respect to the $x=0$ plane, represented by the dashed line in Fig. 1(b). And the width of nanoslit is $w$. The object (a point light source) is located on the left side of the lens at a distance $a$, and the image on the right side at a distance $b$. The plasmonic lens is immersed in the dielectric liquid with permittivity $\varepsilon_d$, and each nanoslit could be also filled with the dielectric liquid. All the components are treated as semi-infinite in the y direction. The Drude model $\varepsilon_m = \varepsilon_\infty - \omega_p^2/[\omega(\omega+iV_c)]$ is used to describe the permittivity of silver at different frequencies, where $\varepsilon_\infty$ =3.2938, plasma frequency $\omega_p$ =1.3552e16 rad/s and collision frequency $V_c$ =1.9944e14 rad/s [16]. All the materials are assumed to be nonmagnetic so that the magnetic permeability $\mu$ is equal to 1 and only the permittivity $\varepsilon$ has been taken into account.

When transverse magnetic (TM) polarized waves impinge on the entrance surface of the silver slab, Surface Plasmon Polaritons (SPPs) are excited [17]. And SPPs propagate through the nanoslit region with specific waveguide modes until they reach the exit surface where they return into the light mode [17, 18]. It is the diffraction and interference of the Surface Plasmon waves that contribute to the transition from the evanescent waves to the propagating waves in the far field region [19], which is the coupling mechanism for far field super-resolution imaging [6]. Therefore, it is theoretically possible for the plasmonic lens to achieve subwavelength imaging in the far field [10]. For an electromagnetic wave incident on such a plasmonic lens, the phase change of the wave is sensitive to the length [7], width [8, 20], and material inside the slit as it passes through each nanoslit [21]. In previous works, the influence of the length and width of the nanoslits on the phase change have been theoretically investigated and widely used for different purposes. In contrast, in this paper, a tunable plasmonic lens is proposed for subwavelength imaging at different wavelengths by varying the surrounding dielectric material while holding nanoslits' lengths, widths and positions constant.

According to the equal optical length principle, the required phase change of light for the point-to-point imaging of an object localized on the axis $x = 0$ can be obtained by:

$$\Delta\phi(x) = 2n\pi + \Delta\phi(0) + \frac{2\pi n_d}{\lambda}\left(a + b - \sqrt{a^2 + x^2} - \sqrt{b^2 + x^2}\right) \quad (1)$$

where $n$ is an arbitrary integer number, $n_d$ is the refractive index of the surrounding dielectric material, $\lambda$ represents the incident wavelength, and $x$ is the position of each nanoslit. For example, when we choose $\lambda = 730 nm$, $a = 1\mu m$, $b = 1\mu m$ and $n_d = 1.33$ (water, $H_2O$) [22], the required phase change at different $x$ positions calculated by Eq. (1) is shown in Fig. 2(a).

On the other hand, assuming that the width of each nanoslit $w$ is much smaller than the incident wavelength, it is reasonable for just considering the fundamental mode in the nanoslit [5]. The complex propagation constant $\beta$ in the slit can be determined by the equation [8, 23]:

$$\tanh\left(\sqrt{\beta^2 - k_0^2\varepsilon_d}\; w/2\right) = \frac{-\varepsilon_d\sqrt{\beta^2 - k_0^2\varepsilon_m}}{\varepsilon_m\sqrt{\beta^2 - k_0^2\varepsilon_d}} \quad (2)$$

where $k_0$ is the wave vector of free space light, $\varepsilon_d$ and $\varepsilon_m$ represent the permittivity of the dielectric inside the nanoslit and metal, respectively. It is clearly seen from this equation that the propagation constant $\beta$ changes as the slit width $w$ varies, when $k_0$, $\varepsilon_d$, $\varepsilon_m$ are fixed. The real and imaginary parts of $\beta$ determine the phase velocity and propagation loss, respectively, of SPPs in the nanoslit, and the phase retardation of light transmitted through the nanoslit can be expressed as:

$$\Delta\phi = \text{Re}(\beta d) + \theta \qquad (3)$$

where $d$ is the thickness of the plasmonic slab, and $\theta$ originates from the multiple reflections between the entrance and exit surface of the slab. Both physical analysis and numerical simulations show that the phase retardation is dominantly determined by the real part of $\beta$ [8]. Therefore, $\Delta\phi$ can be approximated as $\text{Re}(\beta d)$, and it could be obtained from Eq. (2) and (3) that the phase retardation can be tuned by changing the slit width when other parameters are fixed [8-10]. According to Eq. (2) and (3), if $\lambda = 730nm$, $\varepsilon_d = 1.769$ ($\varepsilon_d = n_d^2$, $H_2O$) and $d = 200nm$, the width of each nanoslit at position $x$ could be designed to meet the requirement of phase distribution depicted in Fig. 2(a), as shown in Fig. 2(b).

However, as the incident wavelength changes while other parameters are fixed, the required phase change calculated by Eq. (1) varies; the permittivity of metal $\varepsilon_m$ changes, resulting in the change of propagation constant $\beta$ in each slit according to Eq. (2). Therefore, the point-to-point subwavelength imaging could not be achieved again. There are two ways to solve this problem: First, we could change the positions and widths of the nanoslits to fulfill the requirement of phase change as the incident wavelength varies. Obviously, this is unpractical because we should use different plasmonic lenses to achieve subwavelength imaging at different wavelengths. Secondly, it is possible for subwavelength imaging at different wavelengths by simply varying the surrounding dielectric material. The only requirement for the second method is that we should choose different dielectric fluids ($n_d$) to keep the effective wavelength $\lambda_{eff} = \lambda/n_d$ invariable as the incident wavelength $\lambda$ changes.

On the one hand, it is clearly seen in Eq. (1) that if the effective wavelength $\lambda/n_d$ is constant while the incident wavelength $\lambda$ changes, the required phase change at position $x$ for the point-to-point imaging will be invariable, as shown in Fig. 2(a). In other words, we could realize subwavelength imaging of the same object at the same position as long as $\lambda/n_d$ is invariable. On the other hand, when the position and width of nanoslit is fixed, the phase retardation of light transmitted through the nanoslit $\Delta\phi = \text{Re}(\beta d)$ should be also invariable at different incident wavelengths to accord with the unchanged phase distribution shown in Fig. 2(a). Fortunately, in Eq. (2), as the dielectric material ($\varepsilon_d = n_d^2$) varies to fulfill the requirement of the invariable effective wavelength $\lambda/n_d$ at different incident wavelengths, the propagation constant $\beta$ in each nanoslit changes so little that it could be ignored. And the phase retardation of light transmitted through the nanoslit $\Delta\phi = \text{Re}(\beta d)$ could be also maintained at different wavelengths. Therefore both the required phase change and phase retardation are invariable at different incident wavelengths, resulting in that the imaging process will be slightly influenced by the wavelength variability.

For example, we choose another two different incident wavelengths at $810\,nm$ and $920\,nm$, with corresponding permittivities of the dielectric fluids at 2.200 (carbon tetrachloride, $CCl_4$) and 2.846 (phosphorus tribromide, $PBr_3$) [22], respectively. Other parameters are the same as mentioned above. According to Eq. (2), it could be observed in Fig. 3(a) that the phase retardation of light transmitted through each nanoslit deviate slightly from one another at three different incident wavelengths. The thickness of the metallic wall between two adjacent nanoslits should be larger than the skin depth in metal to prevent the coupling of SPPs during the propagation process. Therefore, the width of the nanoslit should not be large and the maximum value of the width is set at $60\,nm$. When the positions and widths of nanoslits are fixed as such in Fig. 2(b), the calculated phase retardations of the same arrayed nanoslits in the plasmonic lens at three different wavelengths agree well with one another, as shown in Fig. 3(b).

3. **Numerical Simulation and discussion**

Numerical calculations were carried out by using finite-difference time-domain (FDTD) method to illustrate the validity of the tunable plasmonic lens. The simulation dimension is $3\times3\,\mu m^2$ with a grid size of $2\,nm$. The point light sources, which are positioned at the point $x=0$, $z=0.3\mu m$, should be small enough and they are designated to be $\lambda/10$ at different incident wavelengths. The plasmonic lens is located between the plane at $z=1.3\mu m$ and $z=1.5\mu m$. Around the simulation region is the perfect matched layer (PML) boundary condition. As mentioned above, the incident wavelengths are 730nm, 810nm and 920nm, and the corresponding permittivities of the dielectric materials are 1.769, 2.200 and 2.846, respectively. The other parameters are the same as before. The calculated optical field distribution of the simulation result is shown at the left side of Fig. 4(a), (b) and (c). The cross section of image plane at $z=2.43\mu m$ is given at the right side of Fig. 4(a), (b) and (c).

From the left side of Fig. 4(a), (b) and (c), it can be seen that the optical intensity distributions exhibit similarity to one another at three different wavelengths. And the full width at half maximum (FWHM) of the image spots at three different wavelengths are all equal to $345\,nm$ ($0.37\lambda \sim 0.47\lambda$), as shown at the right side of Fig. 4(a), (b) and (c). This is reasonable because the required phase change and phase retardations of the same arrayed nanoslits are invariable at three different wavelengths.

Therefore, the imaging process is slightly influenced by the wavelength variability and FWHM of image spots are the same as one another. The only difference is that the intensity value becomes larger as the incident wavelength increases. The reason is that as the incident wavelength $\lambda$ increases the point light source ($\lambda/10$) becomes larger, resulting in the enhancement of the optical field intensity. Furthermore, the image spot is supposed to be located at point $x=0$, $z=2.5\,\mu m$ because the designed image distance is $b=1\,\mu m$. The slight focal shift of about $70\,nm$ mainly originates from the discreteness of nanoslits and coupling effect of SPPs from the adjacent nanoslits.

## 4. Conclusion

In summary, based on the combination of the particular propagation property of SPPs in nanoslit waveguides and microfluidics, we theoretically proposed a tunable planar plasmonic lens for realizing subwavelength imaging in the far field with different incident wavelengths simply by changing the surrounding dielectric fluid. The numerical simulation performed by FDTD method agrees well with the theoretical analysis, and it is shown that the imaging process is only slightly influenced by the change of incident wavelengths. It is believed that the tunable plasmonic lens could be easily extended to three dimensional and would have novel applications in nanophotonics [24], nanolithography [25, 26], and optical-based on-chip analysis [15], etc.

**Acknowledgment**

This research is supported by 973 Program of China (No.2006CB302900) and Chinese Nature Science Grant (No.60778018).

**Figure and caption**

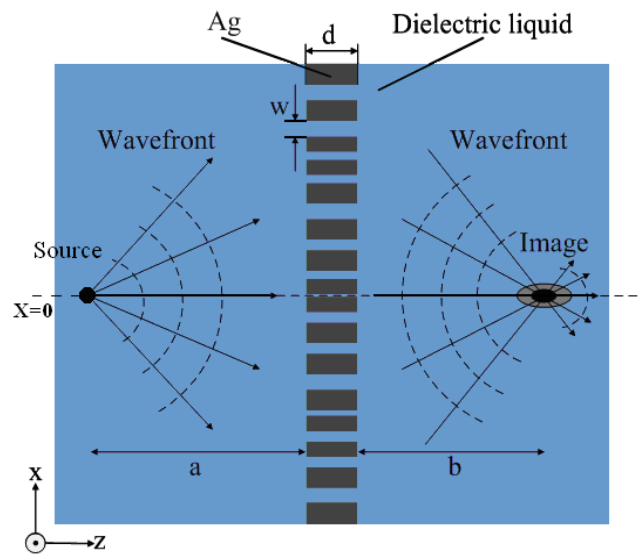

Fig. 1. Schematic drawing of optical imaging process by the tunable optofluidic plasmonic lens.

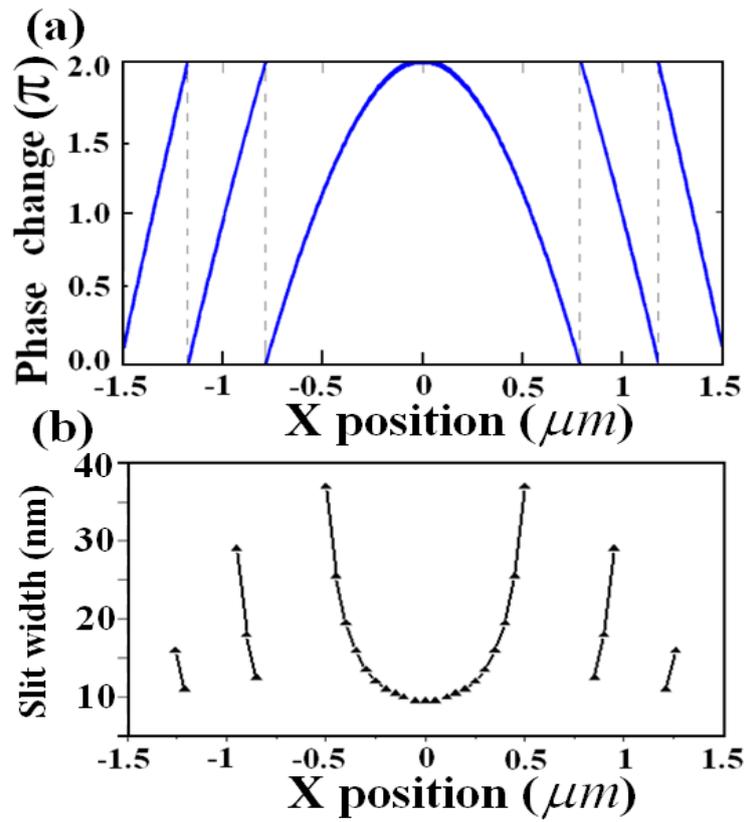

Fig. 2. (a) Required phase change for the point-to-point imaging of an object localized on the axis $x = 0$ as a function of $x$, which is the position of each nanoslit. (b) Distribution of widths and positions of nanoslits in the designed plasmonic lens.

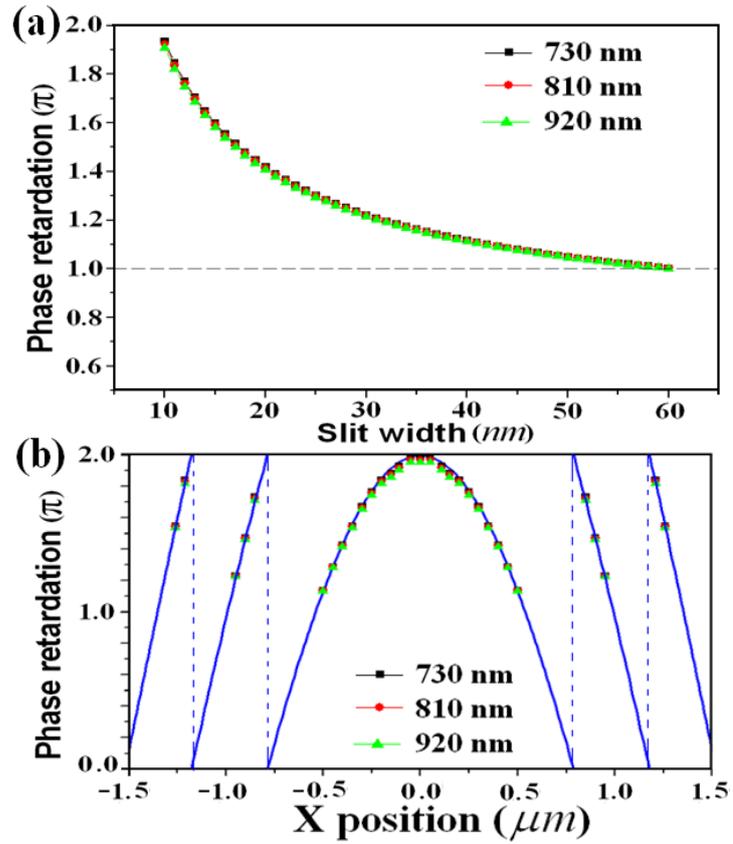

Fig. 3. (a) Dependence of the phase retardation on the slit width at three different wavelengths. The incident wavelengths are $730\,nm$, $810\,nm$ and $920\,nm$. (b) Calculated phase retardations of the arrayed nanoslits in the plasmonic lens at three different wavelengths when the widths and positions of nanoslits are fixed. The blue curve represents the required phase change for the subwavelength imaging.

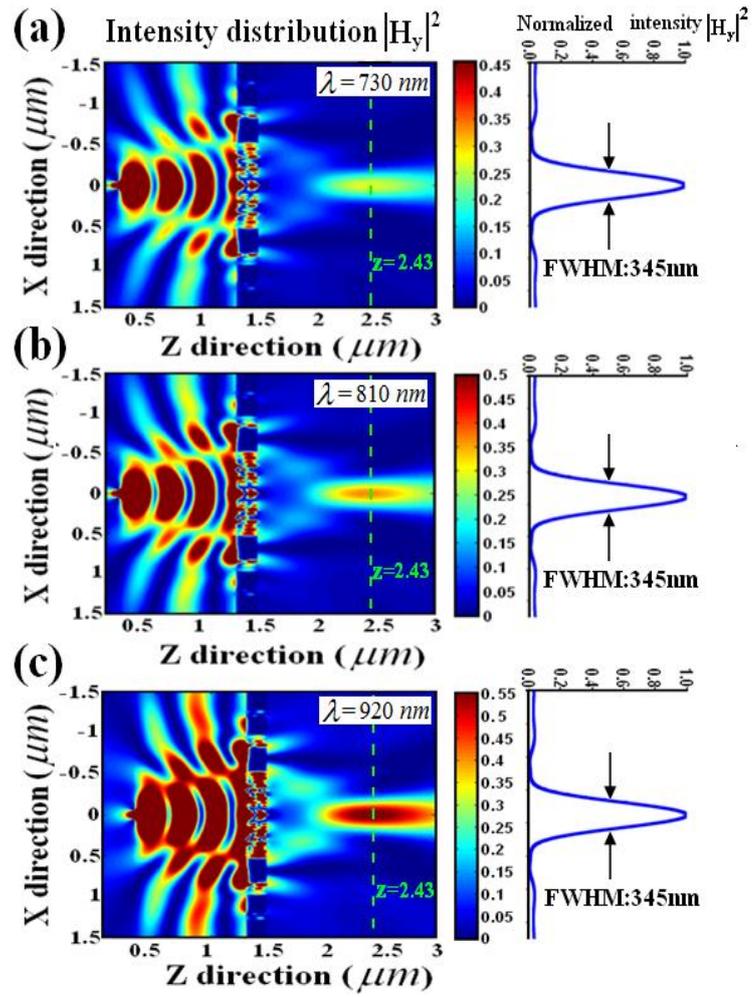

Fig. 4. Calculated optical field distribution of the simulation result (left) and cross section of image plane at $z = 2.43 \mu m$ (right) is shown at wavelengths (a) $730\,nm$, (b) $810\,nm$ and (c) $920\,nm$. The light source is localized at point $x = 0, z = 0.3 \mu m$, and the plasmonic lens ranges from $z = 1.3 \mu m$ to $z = 1.5 \mu m$.